\documentclass[fleqn,10pt]{wlscirep}
\usepackage{pifont}
\usepackage[utf8]{inputenc}
\usepackage[T1]{fontenc}
\usepackage{lineno}

\newcommand{\yes}{\checkmark}
\newcommand{\no}{\ding{55}}
\newcommand{\fn}[1]{\texttt{#1}}
\newcommand{\vn}[1]{\texttt{#1}}

\title{A dataset of mentorship in science with semantic and demographic estimations}

\author[1,*]{Qing Ke}
\author[1]{Lizhen Liang}
\author[2]{Ying Ding}
\author[3]{Stephen V. David}
\author[1,*]{Daniel E. Acuna}
\affil[1]{School of Information Studies, Syracuse University, Syracuse, New York 13244, US}
\affil[2]{School of Information, University of Texas, Austin}
\affil[3]{Oregon Hearing Research Center, Oregon Health and Science University, Portland, Oregon 97239, US}
\affil[*]{Corresponding authors: Qing Ke (qke100@syr.edu) and Daniel E. Acuna (deacuna@syr.edu)}

\begin{abstract}
Mentorship in science is crucial for topic choice, career decisions, and the success of mentees and mentors. Typically, researchers who study mentorship use article co-authorship and doctoral dissertation datasets. However, available datasets of this type focus on narrow selections of fields and miss out on early career and non-publication-related interactions. Here, we describe \textsc{Mentorship}, a crowdsourced dataset of $743\,176$ mentorship relationships among $738\,989$ scientists across 112 fields that avoids these shortcomings. We enrich the scientists' profiles with publication data from the Microsoft Academic Graph and ``semantic'' representations of research using deep learning content analysis. Because gender and race have become critical dimensions when analyzing mentorship and disparities in science, we also provide estimations of these factors. We perform extensive validations of the profile--publication matching, semantic content, and demographic inferences. We anticipate this dataset will spur the study of mentorship in science and deepen our understanding of its role in scientists' career outcomes.
\end{abstract}

\begin{document}

\flushbottom
\maketitle

\thispagestyle{empty}

\section*{Background \& Summary}

Mentorship is a form of guidance provided by a more experienced person (mentor) to a less seasoned one (mentee). Likewise, mentors in science draw from their experiences to help mentees---who often are early-career researchers---navigate various issues inside and outside of academia. Mentorship is a crucial phase in a scientist's development that has long-term effects throughout her career. Mentorship can occur formally through doctoral and postdoctoral advisor--advisee relationships or informally through collaborations. Mentees not only learn new knowledge and skills from mentors but also get involved in mentors' social connections~\cite{austin2002preparing}. Numerous studies have pointed out the association between mentor's characteristics and mentee's academic success, like productivity~\cite{long1985effects, paglis2006does, baruffaldi2016productivity}, career preference and placement~\cite{long1985effects, godechot2016chance, pinheiro2017take}, mentorship fecundity~\cite{malmgren2010role, lienard2018intellectual}, and impact~\cite{Ma14077}. Despite the large role of mentorship and interest in studying it, previous studies have relied on single-field datasets and indirect signals of mentorship (e.g., co-authorship) and therefore have limited generalizability. Large, curated, and open datasets on mentorship have the potential of bringing significant benefit to our understanding of the phenomenon, similar to how citation and publication datasets have accelerated the emerging field of science of science~\cite{Fortunatoeaao0185, Zeng-scisci-2017}.

Studying mentorship requires access to a broad set of relationship types, including publication. There are a few data sources for mentorship in science (Table~\ref{tab:comp}); here, we list a handful of them. The Mathematics Genealogy Project (MGP)~\cite{mgp} is an online database for academic genealogy only in mathematics, though more broadly construed to include ``mathematics education, statistics, computer science, or operations research''. MGP lacks publication records. The Astronomy Genealogy Project is a similar online database confined to astronomy that also does not have publication information~\cite{agp, tenn2016intro}. ProQuest is a database of theses and dissertations predominantly from the US~\cite{proquest}. Although it is multi-disciplinary, it does not disambiguate researchers, making it hard to link advisor and advisee and construct lineages. Also, it does not provide publication information. More importantly, ProQuest is not publicly available, and its access is rate-limited. Apart from genealogy and thesis data, other researchers have proposed to use paper co-authorships as indirect signals of mentorship~\cite{wang2010mining}. However, mentorship can start much earlier than publishing works, and it does not necessarily lead to publications~\cite{acuna2020some}. To summarize, datasets about mentorship in science are in general fragmented.

Here, we start from the Academic Family Tree (AFT) website~\cite{aft} and extend it to create a large-scale dataset of mentorship relationships in science. The AFT is an online portal for mentorship in science. We match each AFT profile to the Microsoft Academic Graph (MAG), a leading bibliographic database~\cite{Wang2019MAG}. Moreover, we apply natural language processing techniques to extract semantic representations of researchers based on deep learning content analysis of their publications. Given the recent interest to understand the role of gender and race/ethnicity in science~\cite{Huang4609}, we also provide estimations of researchers' demographics. Compared to existing databases, our dataset, \textsc{Mentorship} (\textsc{Mentor}ship with \textsc{S}emantic, \textsc{H}ierarchical, and demograph\textsc{I}c \textsc{P}atterns), covers a wide range of disciplines with a richer set of features, making it ideal for studying generalizable mentorship patterns. We expect it to be the base of future studies covering various aspects of scientific mentorship, including semantic and demographic factors.

\section*{Methods}

\subsection*{Data sources}

The AFT website displays researchers' profile information, like direct academic parents and children and a limited set of publication records in the PubMed. Originally focused on neuroscience~\cite{david2012neurotree}, AFT has been expanding to other areas such as chemistry, engineering, and education. As a crowd-sourcing website, contents on AFT are contributed by registered users. Contributions can be diverse, from adding a new researcher to adding mentors, trainees and collaborators of an existing researcher. Visitors can also indicate whether the website has correctly matched a profile with a publication. Due to the crowd-sourcing nature, researchers on AFT may not be a representative sample of the academic population.

In AFT, the user-contributed data are stored in a database consisting of several tables that are available online~\cite{david2021aft}. These tables are the starting point for the present work. In particular, we use four tables: (1) the \texttt{people} table storing researchers' basic information, including person's ID, name, degree, research area, etc.; (2) the \texttt{connect} table detailing mentorship relationships, including its ID, mentee and mentor person IDs, mentorship type (e.g., PhD, postdoctoral advising), and when and where the mentorship occurred; (3) the \texttt{authorPub} table enumerating researchers and their papers as well as meta data of papers; and (4) the \texttt{locations} table listing institutions and their geolocations.

We use the MAG dataset to find papers of AFT researchers. MAG contains information about papers, authors, journals, conferences, affiliations, and citations. One advantage of MAG is that all entities have been disambiguated and associated with identifiers. This dataset has been used in several recent works for author- and venue-level analyses~\cite{Huang4609, Pengeabb9004}. Here we use four tables in MAG: (1) the \texttt{Affiliations} table that lists institution related information; (2) the \texttt{PaperAuthorAffiliations} table that records the name and the affiliation of each authorship; (3) the \texttt{Authors} table that contains author information including names; and (4) the \texttt{Papers} table that consists of paper-related metadata such as digital object identifier (DOI).

\begin{figure}[ht!]
\centering
\includegraphics[width=\linewidth]{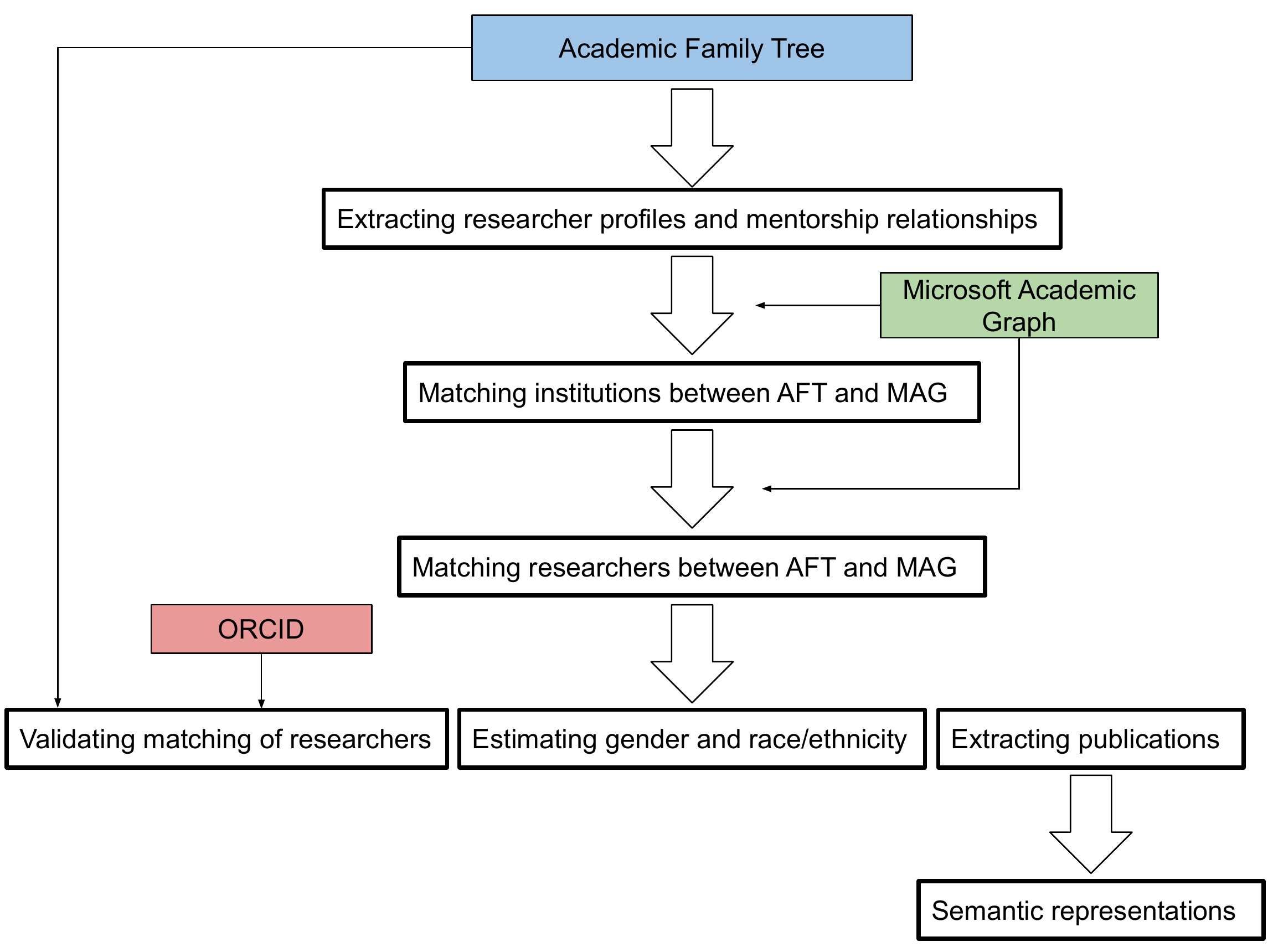}
\caption{Flowchart of the dataset generation process.}
\label{fig:flow}
\end{figure}

Fig.~\ref{fig:flow} provides an overview of how these data sources are used to assemble the dataset presented in our work.

\subsection*{Normalizing researcher profiles}

The \texttt{people} table contains $778\,367$ researchers, uniquely identified by person IDs. We clean this table by ignoring (1) researchers without a first name or last name; (2) researchers who have the same name, institution, and major research area but different IDs as they are likely duplicates; and (3) researchers whose first, middle, or last name contain characters that are not likely to appear in a name, such as ``\&'' and ``;''. These steps leave us with $774\,733$ (99.5\%) researchers. 

Besides person IDs that are used internally in AFT, there are about $1\,600$ researchers whose Open Researcher and Contributor ID (ORCID), a persistent identifier to uniquely identify authors~\cite{aboutorcid}, are available. Although this is a small fraction (0.2\%), we use this information for later validation of our methods. This ORCID information needs cleaning before using it as it contains various ``orcid.org'' prefixes (``https://orcid.org/'', ``http://orcid.org/'', and ``orcid.org/'') and wrong format, which are manually corrected. 

\subsection*{Extracting mentor-mentee pairs}

From the \texttt{connect} table, we filter out mentorship pairs where mentee's person ID or mentor's person ID are not present in the curated list of researchers generated in the previous section. We then drop duplicate records and ignore records where the same relationship ID corresponds to a different mentee or mentor's ID. We obtain $743\,176$ mentorship pairs among $738\,989$ researchers. 

\subsection*{Matching institutions between AFT and MAG}

To facilitate matching AFT researchers with MAG authors, we first match institutions. To do so, we generate a list of rules to normalize AFT institution names iteratively. More specifically, we perform a greedy matching where we sequentially select the unmatched AFT institution with the largest number of researchers associated with it. We then apply several rules to normalize the name so that we can find it in the MAG institution list (see Table~\ref{tab:rule} for the rules). For institutions that cannot be matched using these rules, we manually search them in the MAG if they have at least 200 researchers and discard the remaining institutions. These steps are iterated until no more matches are possible.

\subsection*{Linking AFT researchers to MAG authors}

As described before, one unique feature of our dataset is that we provide lists of publications authored by AFT researchers. One motivation behind this is to access the entire co-authorship network of researchers and potentially understand the topics, venues, and citation dynamics of this network. While AFT already has publication information, it is limited to PubMed only. By matching to MAG, we can access all research areas that are not limited to biomedicine. 

There are two main strategies we follow to find matches. One approach is to find, for each mentor-mentee pair, the list of MAG papers where both of their names appear as co-authors. The other strategy is to match AFT researchers using their names and affiliation information. This second strategy is necessary because some mentees have not published a paper with a mentor yet.

We first elaborate on the first strategy: matching by co-authorship. This strategy involves the following three steps:
\begin{enumerate}
\item First, we prepare a list of mentor-mentee name pairs. To do so, for each AFT researcher, we consider her full name as presented in the AFT. If the first name has more than one character (i.e., not first initial), we also consider two possible variations: (1) first name, middle initial, last name; and (2) first name and last name. For a mentor-mentee pair, we then enumerate all possible name pairs.
\item Second, we scan the MAG to collect papers where the name pair of two co-authors appear in the list of name pairs prepared in the first step. Specifically, for a MAG paper, we collect its co-author names from the \texttt{PaperAuthorAffiliations} and \texttt{Authors} tables. Then, we use the \texttt{nameparser} Python library~\cite{nameparser} to parse a full name into first, middle, and last name. (Author names in the MAG are given as single text.) Next, we consider all possible name pairs of two co-authors and check if each pair is presented in the list of AFT name pairs prepared in the first step. Note that we only consider conference papers, journal articles, and unknown when performing the matching, ignoring the other five types of documents presented in MAG: book chapter, book, dataset, patent, and repository.
\item After scanning the MAG, we obtain a list of associated papers and the MAG author IDs for the mentor and the mentee for each mentor-mentee pair. In total, $359\,238$ AFT researchers have MAG papers associated with them and have at least one corresponding MAG author ID. Among these researchers, $295\,630$ (82.3\%) have only one MAG author ID. For the rest, although multiple MAG ids are associated with them, only one of the ids accounts for more than half of the published works for the vast majority of those researchers. Therefore, we assign the most common MAG author ID to an AFT researcher if there is a single majority (98\% of cases). We drop the remaining 2\% and result in a total of $353\,377$ AFT researchers linked to MAG using co-authorship-based matching.
\end{enumerate}

Next, we match the remaining $421\,356$ unmatched researchers with MAG using their name and institution information. The procedure is similar to co-authorship-based matching. First, we collect, for an AFT researcher, all possible name-institution pairs, by considering her name variations and institutions presented in the profile and mentorship tables (Fig.~\ref{fig:flow}). We then aggregate those pairs across all researchers. Note that for only 928 (0.2\%) unmatched researchers, their name-institution pairs are not unique. Next, we scan the MAG to find papers where the co-authors' name-institution pairs are in the prepared list of name-institution pairs. Through this way, we additionally match $141\,078$ researchers, with the total matched researchers reaching to $494\,455$ ($63.8\%$).

\subsection*{Estimating semantic representations}

Our efforts so far have yielded a list of papers for each AFT researcher who we can match in MAG. Next, we use the titles and abstracts of these papers to construct vector representations of the researcher. Such models can capture semantics, allowing us to apply them in a wide range of scenarios such as comparing the \emph{content} between researchers~\cite{lienard2018intellectual}, recommendation~\cite{achakulvisut2016science}, and matchmaking of scientists~\cite{achakulvisut2020neuromatch}. Here we provide two types of representations; one is based on standard term frequency-inverse document frequency (TF-IDF) vectors, and the other is based on modern deep learning embeddings.

\textbf{TF-IDF representation:} The subset of researchers who we can match in MAG published a total of $16\,942\,415$ papers in MAG. We concatenate the titles and abstracts of these papers. Then using \texttt{scikit-learn}~\cite{scikit-learn}, we preprocess the concatenated text by removing English stop words as well as words appearing only once and apply the TF-IDF transformation. This preprocessing results in a $16\,942\,415 \times 2\,275\,293$ sparse matrix, with each row corresponding to a paper and each column a term. The vector of a researcher is the centroid (average) of the TF-IDF vectors of her documents.

\textbf{Deep learning embedding:} We employ SPECTER~\cite{cohan2020specter}, a representation learning algorithm for scientific documents, to obtain dense vector representations of papers. We concatenate titles and abstracts and use the implementation reported in~\cite{specter}. Each article is represented by a dense vector of 768 dimensions, resulting in a dense $16\,942\,415 \times 768$ matrix for all documents. The vector of a researcher, again, is the average of the vectors of her papers.

\subsection*{Estimating gender and race/ethnicity}

Gender in science has become an important subject of study~\cite{Huang4609}. Here we provide researchers' gender information inferred from their first names. To do so, we encode the character sequence using both the full string and sub-word tokenization as created by a pre-trained BERT model~\cite{devlin2018bert, wolf-etal-2020-transformers}. The output of the BERT model is passed through a pooling layer which creates a vector of 768 elements. This vector is then passed through a dropout layer and softmax layer to produce the final gender predictions. We have three genders in our dataset, two legal labels (female and male) and one unknown label, which attempts to capture potentially non-binary genders. For the training data, we use a combination of datasets. One dataset provides predicted gender of author names in the Author-ity 2009 dataset using the Genni and SexMac tools~\cite{torvik2018authority}. We only maintain data points where Genni and SexMac agree with each other. This filtering step left us with $2\,793\,982$ labeled data points. Another dataset for training comes from the Social Security Administration (SSA) and is about popular newborn names and their gender~\cite{ssa}. The SSA dataset contained $95\,026$ names labeled as ``male'' and ``female''. To reduce the generalization error, we sample each class from the aggregated dataset and obtain a relatively balanced dataset with $1\,500\,000$ data points (male: 600000, female: 600000, unknown: 300000). When training, we sample each of all three labels equally. We use 80\% for training and 20\% for validating. The classes in both splits are also balanced.

We also provide race/ethnicity information of researchers inferred from their full name using a similar architecture. The deep learning architecture is identical to the one used in the gender prediction above: BERT $\rightarrow$ Max Pooling $\rightarrow$ Dropout $\rightarrow$ Softmax. We combine two data sources as our training set. The first one contains the predicted ethnicity of authors in the Author-ity 2009 dataset using the Ethnea tool~\cite{torvik2016ethnea}. We map the predicted categories into four groups: Asian, Hispanic, Black, and White using the mapping described in Table~\ref{tab:race-map}. The second dataset consists of name and ethnicity information extracted from personal profiles on Wikipedia~\cite{ambekar2009name}. We map the Wikipedia labels into the same four categories of ethnicity listed before. Finally, we get a dataset with 720000 data points (black: 180000, Asian: 180000, Hispanic: 180000, white: 180000). The training and validation schedule is similar to the one followed for the gender prediction.

Both models are incorporated in our Python package \texttt{demographicx}~\cite{liangjoss}.

\section*{Data Records}

\begin{figure}[ht!]
\centering
\includegraphics[width=\linewidth]{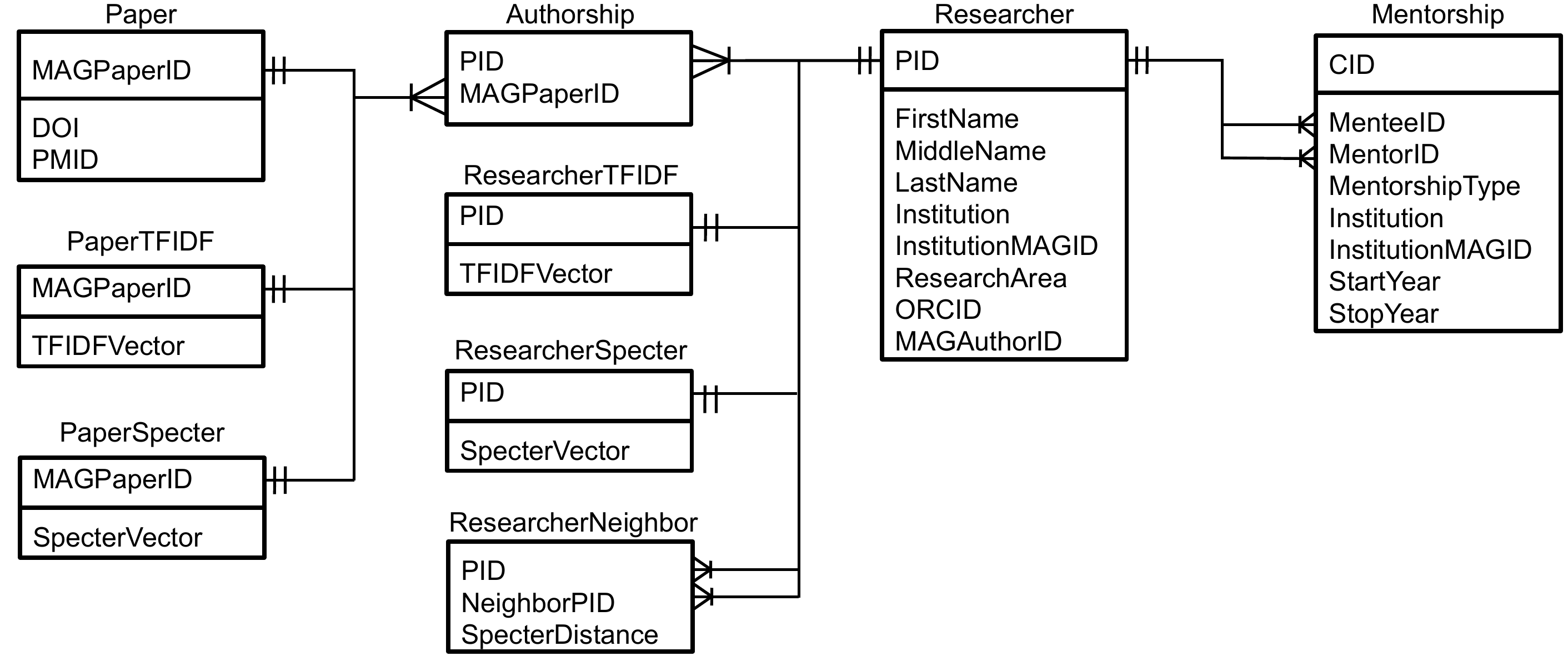}
\caption{Entity-relationship diagram of our dataset.}
\label{fig:er}
\end{figure}

The resulting dataset~\cite{ke2021data} has 9 main tables, shared as the files described below. Fig.~\ref{fig:er} presents the entity-relationship diagram of these tables.

\begin{enumerate}
\item \fn{researcher.csv} is a comma-separated values (CSV) file listing $774\,733$ researchers and contains the following variables: person ID (\vn{PID}), first name, middle name, last name, institution, institution MAG ID, research area, ORCID, and MAG author ID. We also provide an auxiliary file named \fn{first\_name\_gender.csv} that maps first name to inferred gender and an auxiliary file called \fn{full\_name\_race.csv} that maps full name to inferred race/ethnicity.
\item \fn{mentorship.csv} contains mentorship relationships between researchers and has 8 variables: relationship ID (\vn{CID}), mentee's person ID, mentor's person ID, mentorship type, the institution where the mentorship happened, institution MAG ID, and the start year and stop year of the interaction.
\item \fn{authorship.csv} lists all the MAG paper IDs of each researcher and has two columns: person ID (\vn{PID}) and MAG paper ID.
\item \fn{paper.csv} lists 3 types of IDs of each paper: MAG ID, PubMed ID (\vn{PMID}), and DOI.
\item \fn{paper\_tfidf.npz} stores the sparse matrix for paper TF-IDF vectors in Compressed Sparse Row format.
\item \fn{researcher\_tfidf.npz} stores the sparse matrix for researcher TF-IDF vectors in Compressed Sparse Row format.
\item \fn{paper\_specter.pkl} stores SPECTER vectors of papers in the Pickle format.
\item \fn{researcher\_specter.pkl} stores SPECTER vectors of researchers. 
\item \fn{researcher\_neighbor\_specter.csv} list the 9 nearest researchers and the distances to them of each researcher based on SPECTER vectors. It has 3 columns: person ID (\vn{PID}), the neighbor's person ID (\vn{NeighborPID}), and their distance (\vn{SpecterDistance}).
\end{enumerate}

\begin{figure}[ht!]
\centering
\includegraphics[width=\linewidth]{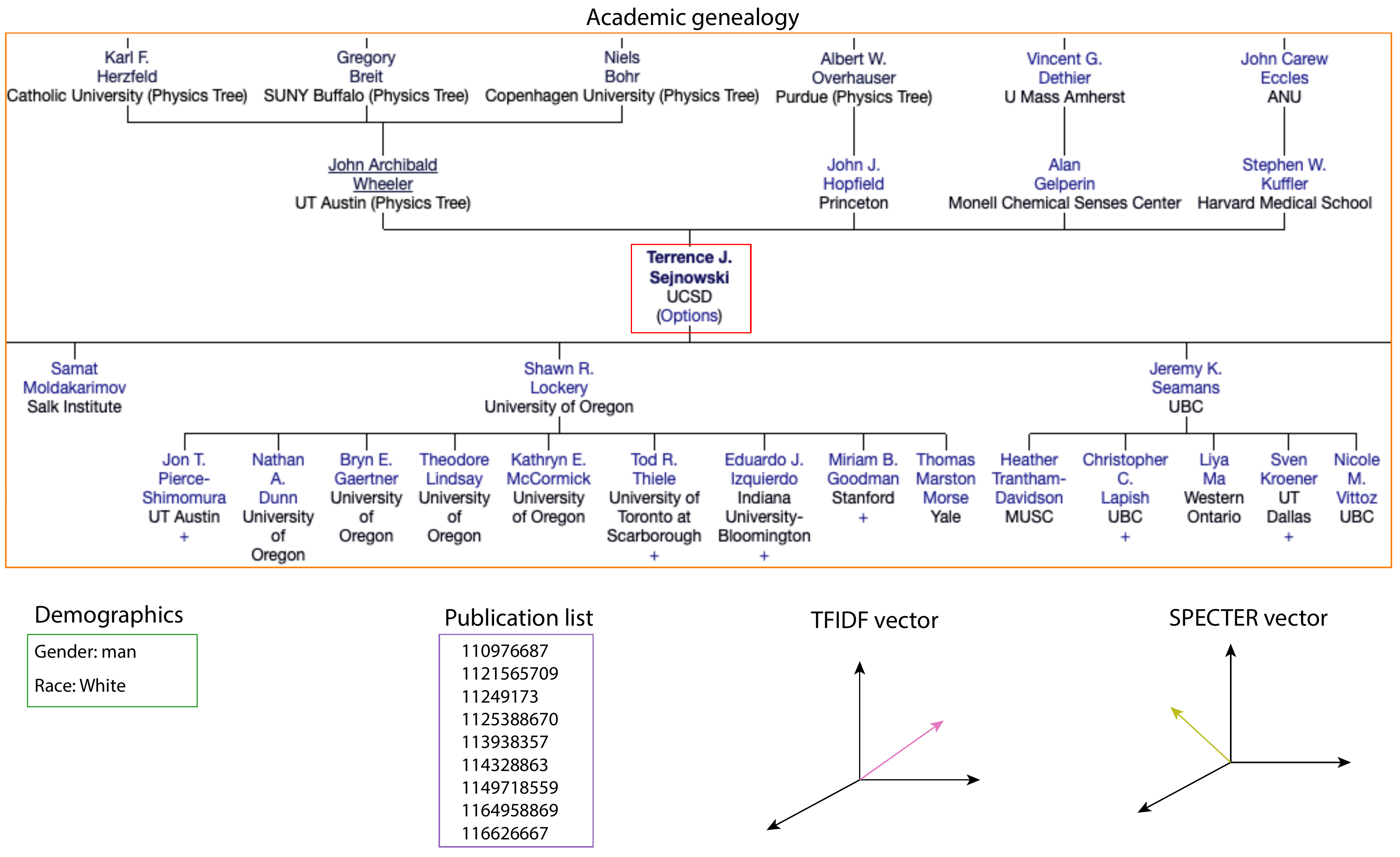}
\caption{Different types of data available for an exemplar researcher (Terrence J. Sejnowski).}
\label{fig:pid76}
\end{figure}

Fig.~\ref{fig:pid76} provides a researcher-centric view of the different types of data available in our dataset.

\section*{Technical Validation}

\subsection*{Validation of gender and ethnicity estimation}

We report in Table~\ref{tab:gender-perf} the performances of our gender prediction algorithm on the validation set and the SSA set. To validate the ``unknown'' class, we used ``unknown'' labels from Authori-ty for names in the SSA dataset labeled ``unknown'' in the Authori-ty dataset. For both sets, our algorithm has good performances for all three categories. Applying the algorithm to our dataset, Table~\ref{tab:gender} presents the numbers of researchers by gender.

Similarly, we test our race/ethnicity prediction algorithm on the validation set and the Wikipedia dataset, obtaining good performances for all four groups (Table~\ref{tab:race-perf}). Table~\ref{tab:race} presents the number of researchers by predicted race/ethnicity using our algorithm.
Even though the model has achieved great performance, we found that African American names are under-represented in the training data set. Since the majority of black names are from outside the U.S., the model made predictions largely based on information about African names outside of the U.S. and might suffer from poor performance when predicting African-American names. Due to the sensitive nature of names and ethnicity, it is hard to find full names of African American names. However, we retrieved 340 names from the Black In Neuro website~\cite{bin} to estimate the extend of the issue. The average probability of predicting a name to be black was 19.5\%, with many names being classified as white names. While names retrieved from Black In Neuro are small and might introduce selection bias, the validation suggests that the ethnicity predictions are poor for African-American names. To improve upon this performance, we created a second model that uses only the surnames reported on the U.S. Census~\cite{us2014frequently}. The performance of this second model was significantly better on the Black in Neuro dataset (30\%). The validation on the U.S. Census reveals that this model has worse performance that the first model above (validation data: Black F1: 0.53, Asian F1: 0.64, Hispanic F1: 0.692, White F1: 0.52). We leave it to the user to determine which of the two models  better serves their analysis.

\subsection*{Validation of mentorship}

Our dataset covers mentorship relationships in multiple disciplines. Table~\ref{tab:area} presents the top 20 most represented areas. Neuroscience is the one with the largest number of researchers, given that AFT was originally aimed for academic genealogy in neuroscience. Social sciences fields, like education, literature, sociology, and economics, are also well represented. Table~\ref{tab:conn-type} gives the count of each type of mentorship.

\subsection*{Validation of linking AFT researchers with MAG authors}

Table~\ref{tab:area} indicates that we can match the majority of researchers in natural sciences, but for social sciences fields like education, literature, we have lower percentages of researchers matched.

To validate our linking of AFT researchers to MAG authors, we take advantage of the fact that their publications are known to be genuinely authored by them for some AFT researchers. With these publications, we examine if they also appear in the publication list of the corresponding matched MAG author. Here we focus on two subsets of AFT researchers: (1) those with papers verified by AFT website users; and (2) those with ORCID available.

Let us describe the first subset. In our previous works~\cite{lienard2018intellectual, david2012neurotree}, we have automatically linked AFT researchers to publications indexed in PubMed. Those matched papers are then displayed on researchers' profile pages. AFT website users who have signed into the website can label whether the authorship is correct. We consider these labeled papers as a validation set to test the performance of our AFT-to-MAG matching of authors. To match these papers to MAG, we rely on their DOIs. For papers without DOI but with PMID, we query PubMed to get their DOI~\cite{pubmed}. 

\begin{figure}[ht!]
\centering
\includegraphics[width=\linewidth]{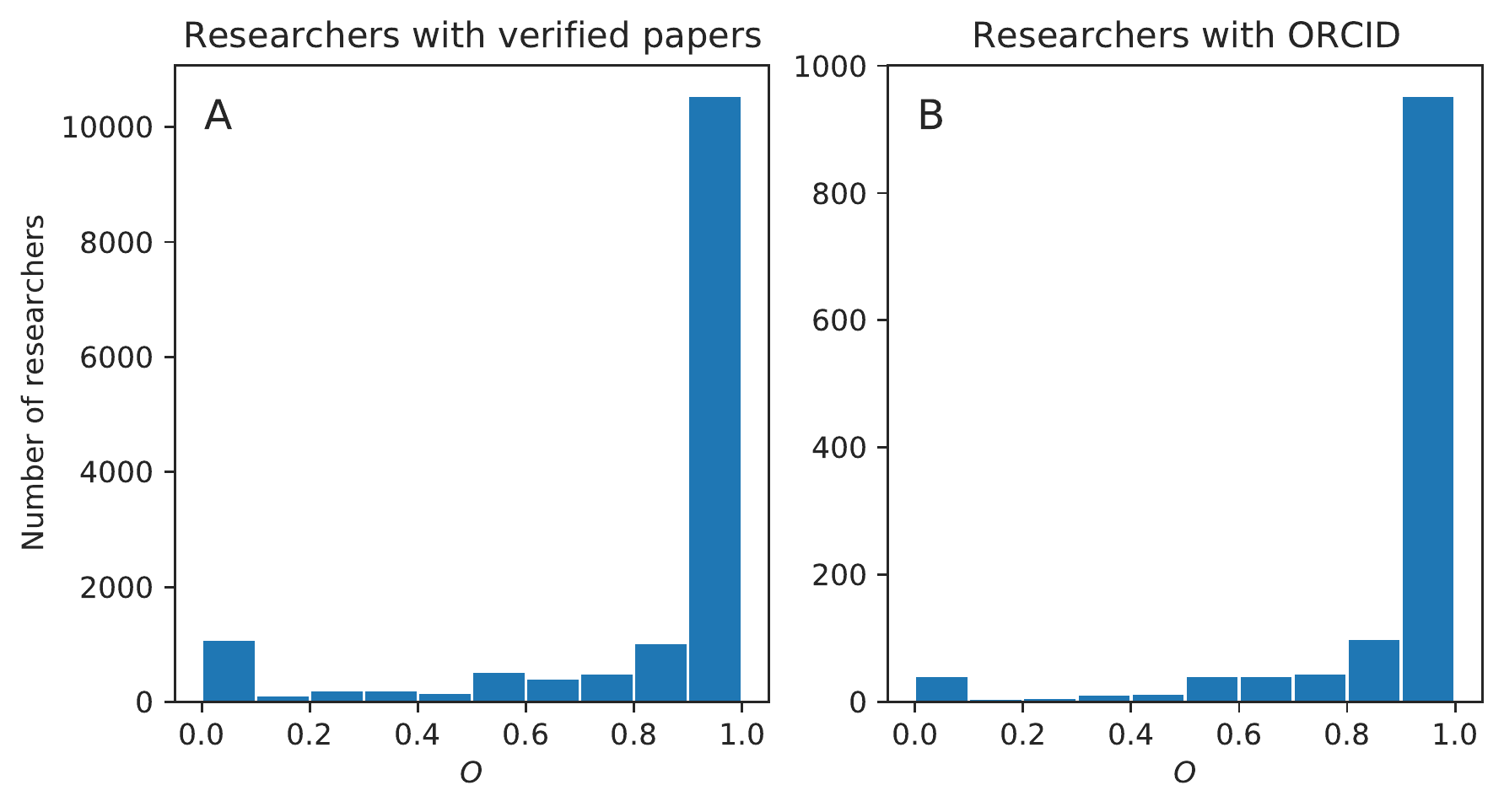}
\caption{Validation of matching AFT researchers with MAG authors. The measure $O$ considers, for an AFT researcher, $a$, the list of papers, $P_a$, genuinely authored by her, and measures the fraction of these papers that also appear in the list of papers of the corresponding matched MAG author. (A) Histogram of $O$ for $14\,824$ researchers with $\left\vert P_a \right\vert > 0$. Here $P_a$ refers to papers that registered AFT website users verify. (B) Histogram of $O$ for $1\,262$ researchers with ORCID identifiers. $P_a$ represents the list of papers extracted from the orcid.org website. In both cases, we observe that for the vast majority of researchers, most of their papers that they genuinely author can be found in the lists of publications of their matched MAG authors, indicating high accuracy of our matching procedure.}
\label{fig:match-val}
\end{figure}

We can now introduce the measure used to quantify the performance of our matching. Let $a$ be an AFT researcher who has at least one verified and $P_a$ the list of her verified papers. Let also $a'$ be the corresponding matched MAG author and $P_{a'}$ the list of papers found on MAG. We calculate the fraction of $P_a$ that appear in $P_a'$, formally:
\begin{equation}
O_a = \frac{\left\vert P_a \cap P_{a'} \right\vert}{\left\vert P_a \right\vert} \, .
\end{equation}
Fig.~\ref{fig:match-val}A, which plots the histogram of $O_a$ for the first subset of researchers, indicates the validity of our matching process; for the vast majority of researchers, we can find most of their verified papers in the publication lists of their matched MAG authors.

Let us describe the second subset: papers listed on the ORCID website ($P_a$). To get these papers, we download the 2019 ORCID Public Data File (the most recent one)~\cite{orcid}, extract documents authored by researchers, and match extracted papers to MAG using their DOI. Fig.~\ref{fig:match-val}B shows the histogram of $O_a$ for the second subset of researchers, indicating most of their papers also appear in publication lists of corresponding matched MAG authors.

\subsection*{Validation of author vector}

\begin{figure}[ht!]
\centering
\includegraphics[width=\linewidth]{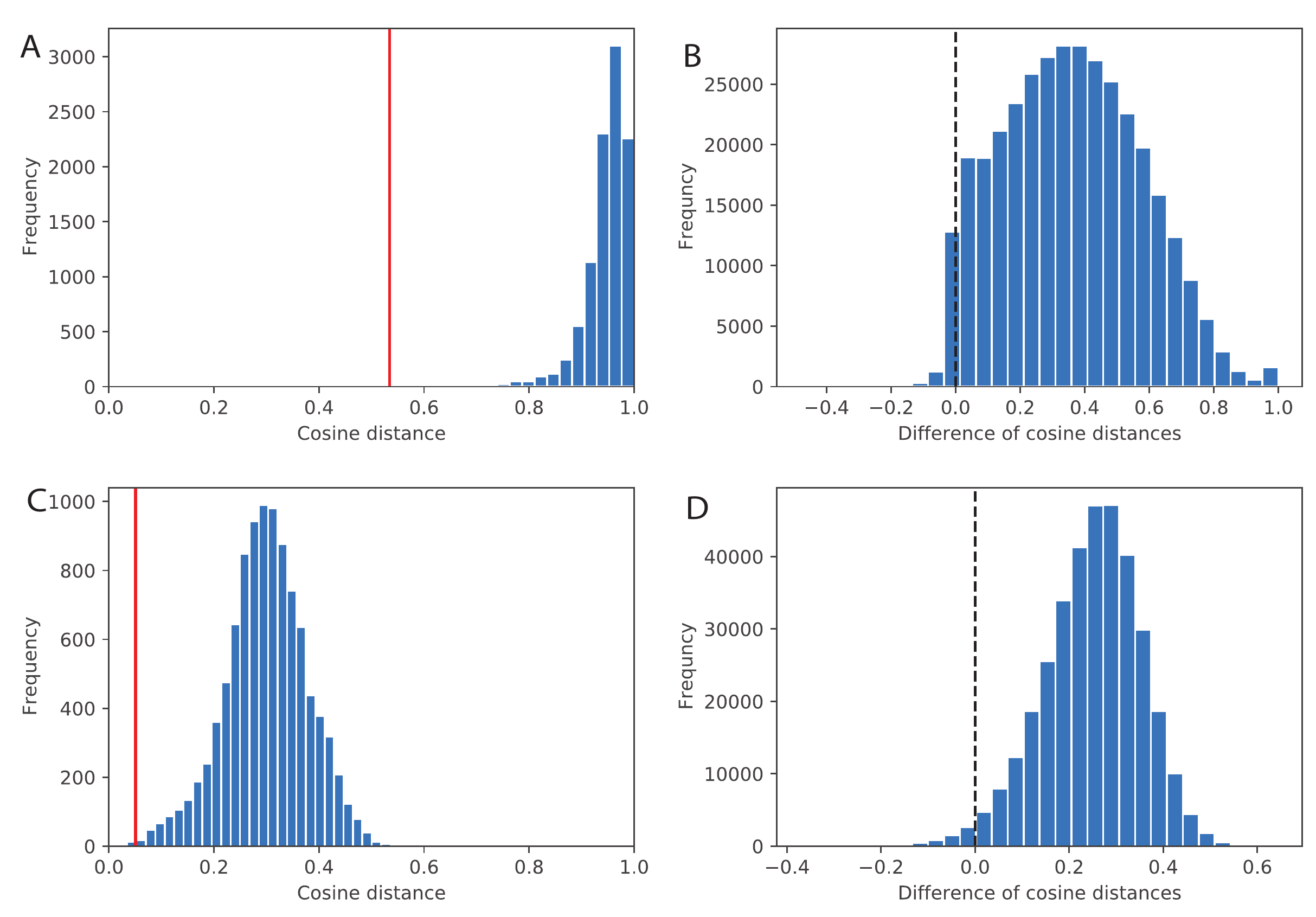}
\caption{Validation of researcher vectors. (A) Histogram of cosine distances of TF-IDF vectors between one researcher $a$ and 10 thousand randomly selected researchers. The red vertical line marks the distance between $a$ and $a$'s Ph.D. mentor, $b$, indicating that $a$ is much closer to her mentor than expected. (B) For each PhD mentee, we calculate the difference of $d(a,c)$ and $d(a,b)$, where $d(a,c)$ is the cosine distance between $a$ and $c$, a randomly selected researcher. The figure shows the histogram of the differences for all Ph.D. mentees, indicating that they are semantically much closer to their mentors than to random researchers for the vast majority of Ph.D. mentees. (C--D) The same as A--B, except that researcher vectors are based on the SPECTER algorithm rather than TF-IDF.}
\label{fig:vec-val}
\end{figure}

We validate researchers' vectors by comparing distances between researchers who belong to different groups. Specifically, in Fig.~\ref{fig:vec-val}A, we show that the cosine distance of the TF-IDF vectors of a particular Ph.D. mentee, $a$, and her mentor, $b$, is much smaller than the distances between $a$ and randomly selected researchers. Generalizing this systematically, for each Ph.D. mentee, we obtain a triplet $(a,b,c)$ where $c$ is a randomly chosen researcher. We then calculate the difference of the distance between $a$ and $c$, $d(a,c)$, and the distance between $a$ and $b$, $d(a,b)$. As we expect, the semantics of a mentee is more similar to her Ph.D. mentor than to a random researcher, and the distance difference is expected to be larger than 0. This pattern is indeed the case for the vast majority (97.4\%) of Ph.D. mentees (Fig.~\ref{fig:vec-val}B). We also replicate these analyses using SPECTER vectors, and the results remain similar (Figs.~\ref{fig:vec-val}C--D): For 98.4\% of Ph.D. mentees, they are semantically closer to their Ph.D. mentors than randomly selected researchers (Fig.~\ref{fig:vec-val}D). The threshold 0 is located at 1.66 and 2.39 standard deviations away from the mean for the TF-IDF case and SPECTER case, respectively, suggesting that SPECTER is a better representation method. 

\begin{figure}[ht!]
\centering
\includegraphics[width=\linewidth]{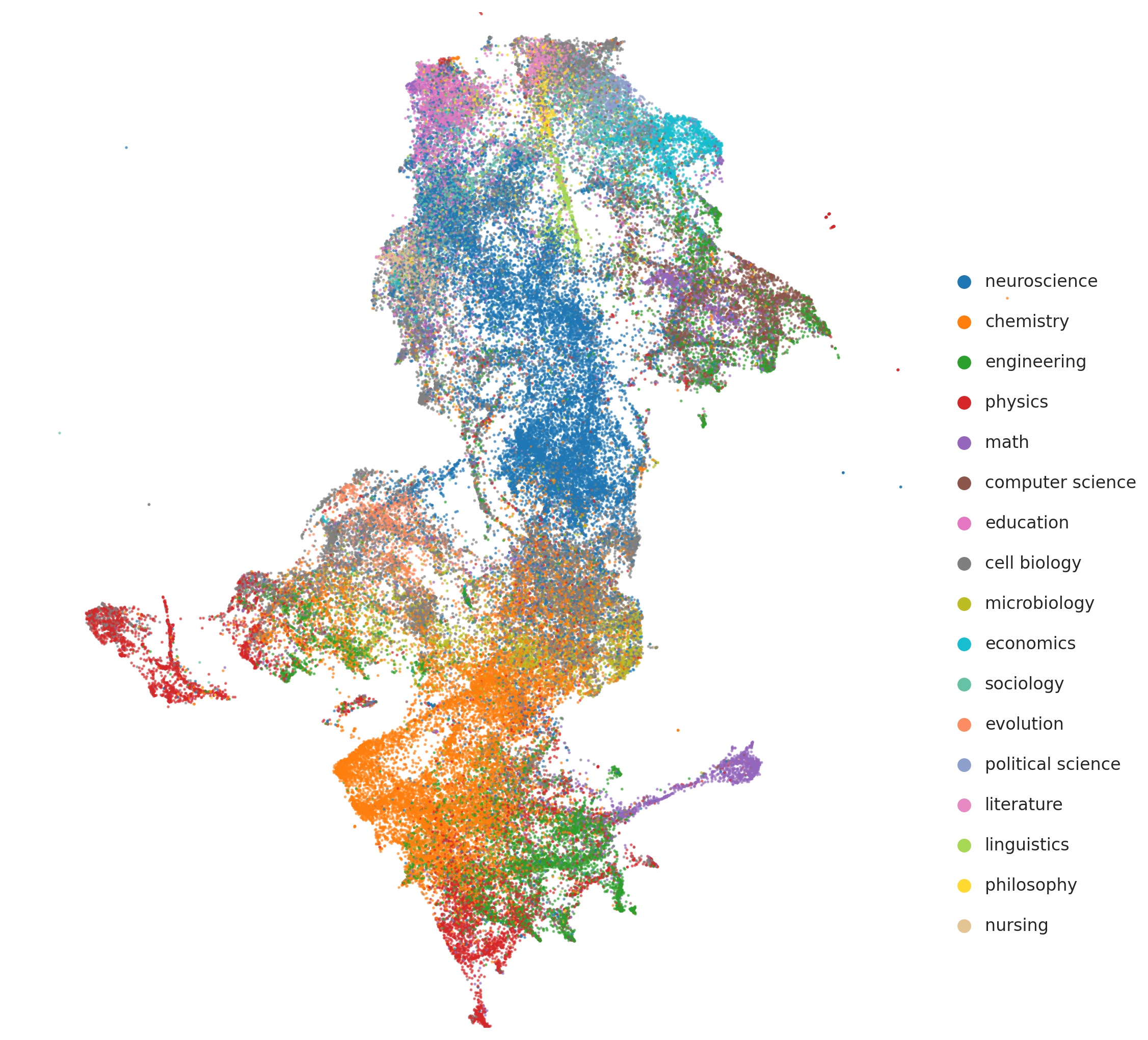}
\caption{The 2-dimensional projections of researchers' SPECTER vectors, obtained using UMAP~\cite{mcinnes2018umap}. The figure shows a 20\% random sample of all researchers. An interactive version can be found at \url{https://scienceofscience.org/mentorship}.}
\label{fig:specter-umap}
\end{figure}

To further show the structure of researchers' SPECTER vectors, we run the UMAP~\cite{mcinnes2018umap} dimension reduction technique to obtain 2-dimensional vectors and display them as a scatter plot for a 20\% random sample of researchers in Fig.~\ref{fig:specter-umap}. As expected, researchers in the same research area are clustered, meaning that they are semantically closer to each other than researchers from other areas. 

\section*{Usage Notes}

Users can integrate our data set with MAG to study the role of mentor in mentee's academic career. MAG provides detailed information about papers and citations, from which users can derive various indicators commonly used in the science of science. We can access MAG data by following the steps outlined on its website~\cite{mag}. In addition to MAG, other identifiers of publications we provide also facilitate integration with other scholarly databases. In particular, users can use CrossRef API to retrieve metadata of papers using DOI~\cite{crossref}. Also, we can use the E-utilities API provided by the National Library of Medicine to obtain metadata of PubMed articles using PMID~\cite{pubmed}.

Users who want to use our released researcher vectors to perform semantic analysis can load the TF-IDF vector file using the SciPy library's \texttt{scipy.sparse.load\_npz} function.

\section*{Code availability}

All the code for generating the dataset and figures is published as IPython notebooks on Github, \url{https://github.com/sciosci/AFT-MAG}. All the coding was completed using Python.


\begin{thebibliography}{10}
\urlstyle{rm}
\expandafter\ifx\csname url\endcsname\relax
  \def\url#1{\texttt{#1}}\fi
\expandafter\ifx\csname urlprefix\endcsname\relax\def\urlprefix{URL }\fi
\expandafter\ifx\csname doiprefix\endcsname\relax\def\doiprefix{DOI: }\fi
\providecommand{\bibinfo}[2]{#2}
\providecommand{\eprint}[2][]{\url{#2}}

\bibitem{austin2002preparing}
\bibinfo{author}{Austin, A.~E.}
\newblock \bibinfo{journal}{\bibinfo{title}{Preparing the next generation of
  faculty: Graduate school as socialization to the academic career}}.
\newblock {\emph{\JournalTitle{The Journal of Higher Education}}}
  \textbf{\bibinfo{volume}{73}}, \bibinfo{pages}{94--122}
  (\bibinfo{year}{2002}).

\bibitem{long1985effects}
\bibinfo{author}{Long, J.~S.} \& \bibinfo{author}{McGinnis, R.}
\newblock \bibinfo{journal}{\bibinfo{title}{The effects of the mentor on the
  academic career}}.
\newblock {\emph{\JournalTitle{Scientometrics}}} \textbf{\bibinfo{volume}{7}},
  \bibinfo{pages}{255--280} (\bibinfo{year}{1985}).

\bibitem{paglis2006does}
\bibinfo{author}{Paglis, L.~L.}, \bibinfo{author}{Green, S.~G.} \&
  \bibinfo{author}{Bauer, T.~N.}
\newblock \bibinfo{journal}{\bibinfo{title}{Does adviser mentoring add value? a
  longitudinal study of mentoring and doctoral student outcomes}}.
\newblock {\emph{\JournalTitle{Research in Higher Education}}}
  \textbf{\bibinfo{volume}{47}}, \bibinfo{pages}{451--476}
  (\bibinfo{year}{2006}).

\bibitem{baruffaldi2016productivity}
\bibinfo{author}{Baruffaldi, S.}, \bibinfo{author}{Visentin, F.} \&
  \bibinfo{author}{Conti, A.}
\newblock \bibinfo{journal}{\bibinfo{title}{The productivity of science \&
  engineering {PhD} students hired from supervisors’ networks}}.
\newblock {\emph{\JournalTitle{Research Policy}}}
  \textbf{\bibinfo{volume}{45}}, \bibinfo{pages}{785--796}
  (\bibinfo{year}{2016}).

\bibitem{godechot2016chance}
\bibinfo{author}{Godechot, O.}
\newblock \bibinfo{journal}{\bibinfo{title}{The chance of influence: A natural
  experiment on the role of social capital in faculty recruitment}}.
\newblock {\emph{\JournalTitle{Social Networks}}}
  \textbf{\bibinfo{volume}{46}}, \bibinfo{pages}{60--75}
  (\bibinfo{year}{2016}).

\bibitem{pinheiro2017take}
\bibinfo{author}{Pinheiro, D.~L.}, \bibinfo{author}{Melkers, J.} \&
  \bibinfo{author}{Newton, S.}
\newblock \bibinfo{journal}{\bibinfo{title}{Take me where i want to go:
  Institutional prestige, advisor sponsorship, and academic career placement
  preferences}}.
\newblock {\emph{\JournalTitle{PLOS ONE}}} \textbf{\bibinfo{volume}{12}},
  \bibinfo{pages}{e0176977} (\bibinfo{year}{2017}).

\bibitem{malmgren2010role}
\bibinfo{author}{Malmgren, R.~D.}, \bibinfo{author}{Ottino, J.~M.} \&
  \bibinfo{author}{Amaral, L. A.~N.}
\newblock \bibinfo{journal}{\bibinfo{title}{The role of mentorship in
  prot{\'e}g{\'e} performance}}.
\newblock {\emph{\JournalTitle{Nature}}} \textbf{\bibinfo{volume}{465}},
  \bibinfo{pages}{622--626} (\bibinfo{year}{2010}).

\bibitem{lienard2018intellectual}
\bibinfo{author}{Li{\'e}nard, J.~F.}, \bibinfo{author}{Achakulvisut, T.},
  \bibinfo{author}{Acuna, D.~E.} \& \bibinfo{author}{David, S.~V.}
\newblock \bibinfo{journal}{\bibinfo{title}{Intellectual synthesis in
  mentorship determines success in academic careers}}.
\newblock {\emph{\JournalTitle{Nature communications}}}
  \textbf{\bibinfo{volume}{9}}, \bibinfo{pages}{1--13} (\bibinfo{year}{2018}).

\bibitem{Ma14077}
\bibinfo{author}{Ma, Y.}, \bibinfo{author}{Mukherjee, S.} \&
  \bibinfo{author}{Uzzi, B.}
\newblock \bibinfo{journal}{\bibinfo{title}{Mentorship and prot{\'e}g{\'e}
  success in stem fields}}.
\newblock {\emph{\JournalTitle{Proceedings of the National Academy of
  Sciences}}} \textbf{\bibinfo{volume}{117}}, \bibinfo{pages}{14077--14083}
  (\bibinfo{year}{2020}).

\bibitem{Fortunatoeaao0185}
\bibinfo{author}{Fortunato, S.} \emph{et~al.}
\newblock \bibinfo{journal}{\bibinfo{title}{Science of science}}.
\newblock {\emph{\JournalTitle{Science}}} \textbf{\bibinfo{volume}{359}},
  \bibinfo{pages}{eaao0185} (\bibinfo{year}{2018}).

\bibitem{Zeng-scisci-2017}
\bibinfo{author}{Zeng, A.} \emph{et~al.}
\newblock \bibinfo{journal}{\bibinfo{title}{The science of science: From the
  perspective of complex systems}}.
\newblock {\emph{\JournalTitle{Physics Reports}}}
  \textbf{\bibinfo{volume}{714-715}}, \bibinfo{pages}{1--73}
  (\bibinfo{year}{2017}).

\bibitem{mgp}
\bibinfo{title}{{Mathematics Genealogy Project}}.
\newblock \bibinfo{howpublished}{\url{https://www.mathgenealogy.org}}
  (\bibinfo{year}{2021}).

\bibitem{agp}
\bibinfo{title}{{Astronomy Genealogy Project}}.
\newblock \bibinfo{howpublished}{\url{https://astrogen.aas.org}}
  (\bibinfo{year}{2021}).

\bibitem{tenn2016intro}
\bibinfo{author}{Tenn, J.~S.}
\newblock \bibinfo{journal}{\bibinfo{title}{Introducing astrogen: The astronomy
  genealogy project}}.
\newblock {\emph{\JournalTitle{Journal of Astronomical History and Heritage}}}
  \textbf{\bibinfo{volume}{19}}, \bibinfo{pages}{298--304}
  (\bibinfo{year}{2016}).

\bibitem{proquest}
\bibinfo{title}{Proquest dissertations \& theses global™}.
\newblock
  \bibinfo{howpublished}{\url{https://about.proquest.com/products-services/pqdtglobal.html}}
  (\bibinfo{year}{2021}).

\bibitem{wang2010mining}
\bibinfo{author}{Wang, C.} \emph{et~al.}
\newblock \bibinfo{title}{Mining advisor-advisee relationships from research
  publication networks}.
\newblock In \emph{\bibinfo{booktitle}{Proceedings of the 16th ACM SIGKDD
  international conference on Knowledge discovery and data mining}},
  \bibinfo{pages}{203--212} (\bibinfo{year}{2010}).

\bibitem{acuna2020some}
\bibinfo{author}{Acuna, D.~E.}
\newblock \bibinfo{journal}{\bibinfo{title}{Some considerations for studying
  gender, mentorship, and scientific impact: commentary on {AlShebli, Makovi,
  and Rahwan (2020)}}}.
\newblock {\emph{\JournalTitle{{OSF Preprints}}}}  (\bibinfo{year}{2020}).

\bibitem{aft}
\bibinfo{title}{{Academic Family Tree}}.
\newblock \bibinfo{howpublished}{\url{https://academictree.org}}
  (\bibinfo{year}{2021}).

\bibitem{Wang2019MAG}
\bibinfo{author}{Wang, K.} \emph{et~al.}
\newblock \bibinfo{journal}{\bibinfo{title}{A review of microsoft academic
  services for science of science studies}}.
\newblock {\emph{\JournalTitle{Frontiers in Big Data}}}
  \textbf{\bibinfo{volume}{2}}, \bibinfo{pages}{45} (\bibinfo{year}{2019}).

\bibitem{Huang4609}
\bibinfo{author}{Huang, J.}, \bibinfo{author}{Gates, A.~J.},
  \bibinfo{author}{Sinatra, R.} \& \bibinfo{author}{Barab{\'a}si, A.-L.}
\newblock \bibinfo{journal}{\bibinfo{title}{Historical comparison of gender
  inequality in scientific careers across countries and disciplines}}.
\newblock {\emph{\JournalTitle{Proceedings of the National Academy of
  Sciences}}} \textbf{\bibinfo{volume}{117}}, \bibinfo{pages}{4609--4616}
  (\bibinfo{year}{2020}).

\bibitem{david2012neurotree}
\bibinfo{author}{David, S.~V.} \& \bibinfo{author}{Hayden, B.~Y.}
\newblock \bibinfo{journal}{\bibinfo{title}{Neurotree: A collaborative,
  graphical database of the academic genealogy of neuroscience}}.
\newblock {\emph{\JournalTitle{PloS one}}} \textbf{\bibinfo{volume}{7}},
  \bibinfo{pages}{e46608} (\bibinfo{year}{2012}).

\bibitem{david2021aft}
\bibinfo{author}{David, S.~V.}
\newblock \bibinfo{title}{Academic family tree data export (version 1.0) [data
  set]. zenodo}.
\newblock \bibinfo{howpublished}{\url{http://doi.org/10.5281/zenodo.4441298}}
  (\bibinfo{year}{2021}).

\bibitem{Pengeabb9004}
\bibinfo{author}{Peng, H.}, \bibinfo{author}{Ke, Q.}, \bibinfo{author}{Budak,
  C.}, \bibinfo{author}{Romero, D.~M.} \& \bibinfo{author}{Ahn, Y.-Y.}
\newblock \bibinfo{journal}{\bibinfo{title}{Neural embeddings of scholarly
  periodicals reveal complex disciplinary organizations}}.
\newblock {\emph{\JournalTitle{Science Advances}}}
  \textbf{\bibinfo{volume}{7}}, \bibinfo{pages}{eabb9004}
  (\bibinfo{year}{2021}).

\bibitem{aboutorcid}
\bibinfo{title}{About {ORCID}}.
\newblock \bibinfo{howpublished}{\url{https://info.orcid.org/what-is-orcid}}
  (\bibinfo{year}{2021}).

\bibitem{nameparser}
\bibinfo{title}{Name parser}.
\newblock
  \bibinfo{howpublished}{\url{https://github.com/derek73/python-nameparser}}
  (\bibinfo{year}{2021}).

\bibitem{achakulvisut2016science}
\bibinfo{author}{Achakulvisut, T.}, \bibinfo{author}{Acuna, D.~E.},
  \bibinfo{author}{Ruangrong, T.} \& \bibinfo{author}{Kording, K.}
\newblock \bibinfo{journal}{\bibinfo{title}{Science concierge: A fast
  content-based recommendation system for scientific publications}}.
\newblock {\emph{\JournalTitle{PloS one}}} \textbf{\bibinfo{volume}{11}},
  \bibinfo{pages}{e0158423} (\bibinfo{year}{2016}).

\bibitem{achakulvisut2020neuromatch}
\bibinfo{author}{Achakulvisut, T.} \emph{et~al.}
\newblock \bibinfo{journal}{\bibinfo{title}{neuromatch: Algorithms to match
  scientists}}.
\newblock {\emph{\JournalTitle{eLife}}}  (\bibinfo{year}{2020}).

\bibitem{scikit-learn}
\bibinfo{author}{Pedregosa, F.} \emph{et~al.}
\newblock \bibinfo{journal}{\bibinfo{title}{Scikit-learn: Machine learning in
  {P}ython}}.
\newblock {\emph{\JournalTitle{Journal of Machine Learning Research}}}
  \textbf{\bibinfo{volume}{12}}, \bibinfo{pages}{2825--2830}
  (\bibinfo{year}{2011}).

\bibitem{cohan2020specter}
\bibinfo{author}{Cohan, A.}, \bibinfo{author}{Feldman, S.},
  \bibinfo{author}{Beltagy, I.}, \bibinfo{author}{Downey, D.} \&
  \bibinfo{author}{Weld, D.~S.}
\newblock \bibinfo{title}{Specter: Document-level representation learning using
  citation-informed transformers}.
\newblock In \emph{\bibinfo{booktitle}{Proceedings of the 58th Annual Meeting
  of the Association for Computational Linguistics}},
  \bibinfo{pages}{2270--2282} (\bibinfo{year}{2020}).

\bibitem{specter}
\bibinfo{howpublished}{\url{https://github.com/allenai/specter}}
  (\bibinfo{year}{2021}).

\bibitem{devlin2018bert}
\bibinfo{author}{Devlin, J.}, \bibinfo{author}{Chang, M.-W.},
  \bibinfo{author}{Lee, K.} \& \bibinfo{author}{Toutanova, K.}
\newblock \bibinfo{title}{{BERT}: Pre-training of deep bidirectional
  transformers for language understanding}.
\newblock In \emph{\bibinfo{booktitle}{Proceedings of the 2019 Conference of
  the North {A}merican Chapter of the Association for Computational
  Linguistics: Human Language Technologies, Volume 1 (Long and Short Papers)}},
  \bibinfo{pages}{4171--4186} (\bibinfo{year}{2019}).

\bibitem{wolf-etal-2020-transformers}
\bibinfo{author}{Wolf, T.} \emph{et~al.}
\newblock \bibinfo{title}{Transformers: State-of-the-art natural language
  processing}.
\newblock In \emph{\bibinfo{booktitle}{Proceedings of the 2020 Conference on
  Empirical Methods in Natural Language Processing: System Demonstrations}},
  \bibinfo{pages}{38--45} (\bibinfo{publisher}{Association for Computational
  Linguistics}, \bibinfo{year}{2020}).

\bibitem{torvik2018authority}
\bibinfo{author}{Torvik, V.}
\newblock \bibinfo{journal}{\bibinfo{title}{{Genni + Ethnea for the Author-ity
  2009 dataset}}}.
\newblock {\emph{\JournalTitle{{University of Illinois at Urbana-Champaign}}}}
  \url{10.13012/B2IDB-9087546_V1} (\bibinfo{year}{2018}).

\bibitem{ssa}
\bibinfo{author}{{Social Security Administration}}.
\newblock \bibinfo{title}{Beyond the top 1000 names}.
\newblock
  \bibinfo{howpublished}{\url{https://www.ssa.gov/oact/babynames/limits.html}}
  (\bibinfo{year}{2021}).

\bibitem{torvik2016ethnea}
\bibinfo{author}{Torvik, V.~I.} \& \bibinfo{author}{Agarwal, S.}
\newblock \bibinfo{title}{Ethnea--an instance-based ethnicity classifier based
  on geo-coded author names in a large-scale bibliographic database}
  (\bibinfo{year}{2016}).

\bibitem{ambekar2009name}
\bibinfo{author}{Ambekar, A.}, \bibinfo{author}{Ward, C.},
  \bibinfo{author}{Mohammed, J.}, \bibinfo{author}{Male, S.} \&
  \bibinfo{author}{Skiena, S.}
\newblock \bibinfo{title}{Name-ethnicity classification from open sources}.
\newblock In \emph{\bibinfo{booktitle}{Proceedings of the 15th ACM SIGKDD
  International Conference on Knowledge Discovery and Data Mining}},
  \bibinfo{pages}{49--58} (\bibinfo{year}{2009}).

\bibitem{liangjoss}
\bibinfo{author}{Liang, L.} \& \bibinfo{author}{Acuna, D.~E.}
\newblock \bibinfo{title}{demographicx: {A} {P}ython package for estimating
  gender and ethnicity using deep learning transformers}.
\newblock \bibinfo{howpublished}{Preprint
  {\url{https://doi.org/10.5281/zenodo.4898367}}} (\bibinfo{year}{2021}).

\bibitem{ke2021data}
\bibinfo{author}{Ke, Q.}, \bibinfo{author}{Liang, L.}, \bibinfo{author}{Ding,
  Y.}, \bibinfo{author}{David, S.~V.} \& \bibinfo{author}{Acuna, D.~E.}
\newblock \bibinfo{title}{A dataset of mentorship in science with semantic and
  demographic estimations}.
\newblock \bibinfo{howpublished}{\emph{zenodo}
  \url{https://doi.org/10.5281/zenodo.4917086}} (\bibinfo{year}{2021}).

\bibitem{bin}
\bibinfo{title}{{Black In Neuro}}.
\newblock
  \bibinfo{howpublished}{\url{https://violin-porcupine-x6rg.squarespace.com/}}
  (\bibinfo{year}{2021}).

\bibitem{us2014frequently}
\bibinfo{author}{Bureau, U.~C.}
\newblock \bibinfo{title}{Frequently occurring surnames from the census 2000}.
\newblock
  \bibinfo{howpublished}{\url{https://www.census.gov/topics/population/genealogy/data/2010_surnames.html}}
  (\bibinfo{year}{2014}).

\bibitem{pubmed}
\bibinfo{title}{Get article metadata}.
\newblock
  \bibinfo{howpublished}{\url{https://www.ncbi.nlm.nih.gov/pmc/tools/get-metadata/}}
  (\bibinfo{year}{2021}).

\bibitem{orcid}
\bibinfo{title}{{ORCID} public data file 2019}.
\newblock \bibinfo{howpublished}{\emph{figshare}
  \url{https://orcid.figshare.com/articles/ORCID_Public_Data_File_2019/9988322/2}}
  (\bibinfo{year}{2020}).

\bibitem{mcinnes2018umap}
\bibinfo{author}{McInnes, L.}, \bibinfo{author}{Healy, J.} \&
  \bibinfo{author}{Melville, J.}
\newblock \bibinfo{journal}{\bibinfo{title}{{UMAP}: Uniform manifold
  approximation and projection for dimension reduction}}.
\newblock {\emph{\JournalTitle{arXiv preprint arXiv:1802.03426}}}
  (\bibinfo{year}{2018}).

\bibitem{mag}
\bibinfo{title}{Get {Microsoft Academic Graph} on {Azure} storage}.
\newblock
  \bibinfo{howpublished}{\url{https://docs.microsoft.com/en-us/academic-services/graph/get-started-setup-provisioning}}
  (\bibinfo{year}{2020}).

\bibitem{crossref}
\bibinfo{title}{Crossref {REST API}}.
\newblock
  \bibinfo{howpublished}{\url{https://github.com/CrossRef/rest-api-doc}}
  (\bibinfo{year}{2021}).

\end{thebibliography}

\section*{Acknowledgements}

This work was partially supported by NSF grant \#1933803. We thank Longfeng Wu for initial data explorations and Leah Schwartz for valuable discussions.

\section*{Author Contributions Statement}

Q.K.\ and D.A.\ conceived the experiments, Q.K.\ conducted the experiments, L.L.\ performed gender and race/ethnicity estimations, Q.K.\ and D.A.\ analysed the results. All authors edited and reviewed the manuscript.

\section*{Competing Interests}

The authors declare no competing interests.

\section*{Figures \& Tables}

\clearpage

\begin{table}[ht!]
\centering
\begin{tabular}{l|l|l|c|c|c}
\hline
Database & Discipline & Country & Tree & Publication data & Open \\
\hline
\textsc{Mentorship} & all & world-wide & \yes & \yes & \yes \\
\hline
Mathematics Genealogy Project & Math & world-wide & \yes & \no & \yes \\ 
\hline
Astronomy Genealogy Project & Astronomy & world-wide & \yes & \no & \yes \\
\hline
ProQuest & all & US & \no & \no & \no \\
\hline
\end{tabular}
\caption{Comparison of existing datasets of mentorship in science with ours (\textsc{Mentorship}).}
\label{tab:comp}
\end{table}

\begin{table}[ht!]
\centering
\begin{tabular}{l}
\hline
replace ``, '' with ``-'' \\
replace ``, '' with `` AT '' then replace ``-'' with `` '' \\
replace `` \& '' with ``\&'' \\
replace ``, '' with `` '' \\
remove ``THE'' then replace `` - '' with ``–'' \\
replace `` AT '' with `` '' \\
replace `` AND '' with `` \& '' \\
extract text in parentheses then replace ``UNIVERSIDAD DE '' (``UNIVERSIDADE DE '') with ``UNIVERSITY OF '' \\
remove accents \\
text before ``,'' \\
replace ``-'' with ``–'' \\
replace ``,'' with `` '' \\
replace `` \& '' with `` AND '' \\
replace `` - '' with `` '' \\
replace `` AND '' with `` '' \\
replace ``IIT '' with ``INDIAN INSTITUTE OF TECHNOLOGY '' \\
\hline
\end{tabular}
\caption{A list of rules to normalize AFT institution names used to match with MAG institutions.}
\label{tab:rule}
\end{table}

\begin{table}[ht!]
\centering
\begin{tabular}{l | l}
\hline
Race for prediction & Race in Ethnea \\
\hline
Asian & Arab, Chinese, Indian, Indonesian, Israeli, Japanese, Korean, Mongolian, Polynesian, Thai, Vietnamese \\
White & Baltic, Dutch, English, French, Greek, German, Hungarian, Italian, Nordic, Romanian, Slav, Turkish \\
Hispanic & Caribbean \\
Black & African \\
\hline
\end{tabular}
\caption{Mapping between race categories in the Ethnea and ours used for prediction.}
\label{tab:race-map}
\end{table}

\begin{table}[ht!]
\centering
\begin{tabular}{l|c c c|c c c}
\hline
         &  \multicolumn{3}{c|}{Validation set} & \multicolumn{3}{c}{SSA names} \\
         & F-1   & accuracy & AUROC & F-1  & accuracy & AUROC \\
\hline
Male    & 0.961 & 0.972    & 0.993 & 0.813 & 0.771    & 0.954 \\
Female  & 0.975 & 0.979    & 0.996 & 0.915 & 0.885    & 0.965 \\
Unknown & 0.889 & 0.862    & 0.966 & 0.504 & 0.664    & 0.860 \\
\hline
\end{tabular}
\caption{Performances of gender prediction.}
\label{tab:gender-perf}
\end{table}

\begin{table}[ht!]
\centering
\begin{tabular}{l|r}
\hline
Gender & \# researchers \\
\hline
male   &  374199 \\
female &  264263 \\
unk    &  135732 \\
\hline
\end{tabular}
\caption{Number of researchers by gender. Here, the gender of the researcher is estimated by an algorithm using their first name. We acknowledge that there could be a great deal of noise and bias in this estimation. However, we believe it is better to open our algorithm to the community instead of analyzing proprietary software that does not publicize data used and performance metrics.}
\label{tab:gender}
\end{table}

\begin{table}[ht!]
\centering
\begin{tabular}{l|c c c|c c c}
\hline
         &  \multicolumn{3}{c|}{Validation set} & \multicolumn{3}{c}{Wikipedia} \\
         & F-1   & accuracy & AUROC & F-1   & accuracy & AUROC \\
\hline
Black    & 0.976 & 0.999    & 0.999 & 0.987 & 0.999    & 0.996 \\
Hispanic & 0.936 & 0.928    & 0.990 & 0.822 & 0.788    & 0.964 \\
White    & 0.907 & 0.902    & 0.983 & 0.850 & 0.856    & 0.963 \\
Asian    & 0.941 & 0.931    & 0.989 & 0.859 & 0.843    & 0.962 \\
\hline
\end{tabular}
\caption{Performances of race/ethnicity prediction.}
\label{tab:race-perf}
\end{table}

\begin{table}[ht!]
\centering
\begin{tabular}{l|r}
\hline
Race/ethnicity & \# researchers \\
\hline
White    &  508923 \\
Asian    &  177649 \\
Hispanic &   68664 \\
Black    &   18958 \\
\hline
\end{tabular}
\caption{Number of researchers by estimated race/ethnicity.}
\label{tab:race}
\end{table}

\begin{table}[ht!]
\centering
\begin{tabular}{lrrrr}
\toprule
            area &      researchers &   \% researchers &      researchers matched &    \% matched \\
\midrule
    neuroscience &           135756 &             16.7 &                    93769 &         69.1 \\
       chemistry &           104450 &             12.9 &                    85585 &         81.9 \\
     engineering &            56898 &              7.0 &                    45004 &         79.1 \\
       education &            56580 &              7.0 &                    17978 &         31.8 \\
         physics &            49582 &              6.1 &                    37714 &         76.1 \\
            math &            35651 &              4.4 &                    22707 &         63.7 \\
      literature &            28257 &              3.5 &                     7449 &         26.4 \\
       sociology &            25453 &              3.1 &                    12618 &         49.6 \\
       economics &            23497 &              2.9 &                    12841 &         54.6 \\
computer science &            22399 &              2.8 &                    18315 &         81.8 \\
    cell biology &            20970 &              2.6 &                    18087 &         86.3 \\
political science&            18914 &              2.3 &                     8654 &         45.8 \\
        theology &            17448 &              2.1 &                     3726 &         21.4 \\
    microbiology &            17230 &              2.1 &                    14759 &         85.7 \\
      phillosopy &            17035 &              2.1 &                     6253 &         36.7 \\
     linguistics &            13952 &              1.7 &                     6685 &         47.9 \\
         nursing &            13825 &              1.7 &                     6207 &         44.9 \\
          phtree &            13637 &              1.7 &                     8986 &         65.9 \\
    anthropology &            13471 &              1.7 &                     6185 &         45.9 \\
       evolution &            13417 &              1.7 &                    10494 &         78.2 \\
\bottomrule
\end{tabular}
\caption{The top 20 most represented major research areas.}
\label{tab:area}
\end{table}

\begin{table}[ht!]
\centering
\begin{tabular}{c|l|r}
\hline
Mentorship type & Definition & Count \\
\hline
0 & Research assistant & 18850 \\ \hline
1 & Graduate student & 630439 \\ \hline
2 & Postdoctoral & 68652 \\ \hline
3 & Research scientist & 7402 \\ \hline
4 & Collaborator & 17833 \\ \hline
\end{tabular}
\caption{Mentorship type definition and statistics.}
\label{tab:conn-type}
\end{table}

\end{document}